# Ambitious forest biodiversity conservation under scarce public funds: introducing a deferrence mechanism to conservation auctions

Johanna Kangas[1,*], Janne S. Kotiaho[2], Markku Ollikainen[1]

[1] Department of Economics and Management, University of Helsinki

[2] Department of Biological and Environmental Science & School of Resource Wisdom, University of Jyväskylä

*Corresponding author: johanna.a.kangas@helsinki.fi

**Abstract**

The European Union's Biodiversity Strategy sets an ambitious goal to increase the area of protected land and sea to 30% with 10% devoted to strict protection by 2030. The large land areas required to fulfil the conservation target and the quick schedule of implementation challenge both the current policy instruments and public funding for conservation. We introduce a deferrence mechanism for forest conservation by using procurement auctions. Deferring the conservation payments allows the government to conserve large areas in a quicker schedule and distributing the financial burden of conservation cost for a longer period of time. The deferred payments are paid an interest. The interest earning and an auction mechanism for downpayments strengthens the incentives for landowners to take part in conservation. We characterize the general properties of the mechanism and run numerical simulations to find that the deferrence mechanism facilitates a quick conservation of stands and thereby minimizes the loss of ecologically valuable sites caused by harvesting risks. The analysis suggests that keeping the lending period no longer than 10 years and paying a 3% interest rate provides a compromise that works rather well and outperforms the up-front mechanism in most cases.

1. Introduction

The decline of biodiversity is alarming all over the world (IPBES 2019). In boreal forests, large amounts of habitats and species are endangered due to intensive commercial management and stress caused by climate change. The European Union's (EU) Biodiversity Strategy aims to halt biodiversity loss by 2030 relying on ambitious goals to increase the area of protected land and sea to 30%, of which one third should be strict protection (European Commission, 2021). The required land areas in boreal and temperate forests are large. As most forests in Europe are owned by private forest landowners, conservation relies mostly on voluntary instruments (Hanley et al. 2012).

Ambitious targets for forest conservation challenges not only current instruments applied to conservation but also government funding. Many European countries are wrestling with insufficient public funds and budget deficits. At the same time many countries face the risk that old-growth and other ecologically valuable forests become harvested, as demand for timber has increased and the forest industry is expanding. Relying on gradual conservation restricted by insufficient annual budgets will not be enough to secure the remaining ecologically valuable forests. Thus, the traditional use of up-front payments, that is, paying full compensation at the time of conservation, faces severe limitations.

The crucial question in the implementation of biodiversity strategy for European union member states is, how to execute rapidly ambitious forest biodiversity conservation when the government conservation budget is limited, and old-growth and other ecologically valuable stands are subject to harvesting and development risks? In this paper, we suggest a novel approach to moderate both challenges and quickly conserve larger areas of forest. The novel instrument we propose is a deferrence mechanism for forest conservation, implemented through competitive auctioning. In this paper, we develop a theory of deferrence auctioning and apply it numerically to large data set of forests.

Deferrence mechanisms are widely used in consumer goods markets. The traditional deferrence mechanism consists of three components: the initial downpayment, instalments per period and the interest rate on the consumption loan (e.g. Maesen and Ang 2024; Siemens 2007). For consumers, this mechanism allows buying now and paying later, which reduces the annual financial burden of consumption.

To date, there are no applications of deferrence mechanism in conservation auctioning. It can be applied in the form of a forest biodiversity auction as follows. The government announces a conservation program for old-growth and other ecologically valuable stands and invites forest landowners to offer their stands to the program. To guide landowners, the government defines the number of periods for annual instalments and the fixed interest rate paid for the loan the forest landowners give to the government. In the usual case of lending money, the lender determines the required interest payment for the loan over the loan period and divides this total sum over the years of the loan contract as requested payments. Instead, in this government bidding system, the interest rate is fixed and forest landowners are invited to offer their stands by submitting their requested initial downpayment as their bids drawing on the commercial value of the stand and their own assessment of their winning probabilities. The difference between the conservation payment (equal to the stand value) and the initial downpayment is the loan that the landowner gives to the government and gets paid for it by instalments, including interest.

We hypothesize that by using this mechanism, the government could be able to conserve large areas now with a bigger initial budget and relatively small subsequent conservation budgets when compared with an up-front payment mechanism that has the same present value for the conservation budget. Thus, it can better safeguard old-growth and other ecologically valuable stands from harvesting. While allowing conservation of large areas now, this mechanism enables distributing the financial burden of conservation costs over a longer period of time. Therefore, it would reduce the degradation of valuable stands and biodiversity loss caused by harvesting and therefore promote biodiversity conservation by speeding up conservation. Incentives matter, however. It is important to note that by omitting the one-time up-front payments, landowners' willingness to participate requires that the interest earning on the loan is high enough and forest landowners can bid the size of the downpayment with which they are willing to participate. This interplay between the interest earning and the downpayment is the focal point of this paper. We examine the feasibility of (reverse) conservation auctioning with a deferrence mechanism and landowners' incentives to participate in it.

Our work is the first of its kind. Mitani and Lindhjem (2021) find that some forms of deferrence mechanisms have also been included in payments for ecosystem services schemes. However, in these schemes, the conservation payment is often simply divided into instalments (Mitani and Lindhjem 2021; Mäntymaa et al. 2018), and the landowner is not compensated for the deferrence of payments by paying interest. Consequently, these payment mechanisms have been found to decrease participation. Following the deferrence mechanism from consumer goods markets, incentives to participate would be better if the producer of ecosystem services commits to produce ecosystem services with an initial downpayment and consequent instalments with some interest, thus giving a temporary loan to the buyer of the ecosystem services. The incentive to commit to deferrence emerges if the stream of rental payments to the loan from the buyer is large enough.

Closest to our work are Sullivan et al. (2005) and Mäntymaa et al. (2018) who analysed deferred payments for temporary conservation agreements for forest banking and for forest landscape conservation, respectively. Sartori et al. (2024) examined rotation deferral in harvests for carbon benefits (Sartori et al. 2024). The use of auction mechanisms for the up-front payments has been analysed in many papers (Juutinen and Ollikainen 2010; Juutinen et al. 2013; Kangas and Ollikainen 2025). These papers focus mostly on the supply side of the conservation schemes. Our work analyses not only supply of sites but also the properties of conservation outcomes if deferred payments are used instead of one-time up-front payments.

## 2. Theoretical framework: extending the theory of reverse auctioning to deferrence for forest conservation

Consider a government deciding to introduce a deferrence mechanism through competitive bidding to forest conservation. To this end, the government announces a conservation program for ecologically valuable stands and invites forest landowners to offer their stands to the program with a bid that consist of their requested initial downpayment. To guide landowners, the government defines a priori the number of periods for annual instalments and the interest rate for the loan, as well as the conservation payment (equal to the net present value of the stand). Thus, forest landowners are invited to offer their stands by submitting bids based on the commercial value of the stand and the requested initial downpayment.

To operationalize the key concepts, let $R$ denote landowner's revenue, which is the total conservation payment from the government to the forest landowner, and $c$ the size of the downpayment. Further, let the interest rate paid for the loan be $r$, the lending period $t$, and the periods of instalments (years) $x$. Commercial timber prices are used to determine the conservation payment, i.e., economic value of the stand, and we denote it by $V_1$. Thus, the difference between the conservation payment and downpayment defines the loan that the landowner gives to the government, $V_1 - c$ for the fixed period of time, $t$. In the deferrence mechanism, the interest earning is defined for the whole loan and the lending period. Then, this amount is divided by payback periods, giving the instalments.

For a fixed interest rate, the interest earning in the mechanism is defined as $(1 + r)^t(V_1 - c)$. Then, we can define $m = \frac{(1+r)^t(V_1-c)}{x}$ as the size of the annual instalment, including the interest on the loan. Note that this expression shows a trade-off between the downpayment and the interest earning, as $\frac{dm}{dc} = -\frac{(1+r)^t}{x} < 0$. The expected revenue to the forest landowner from participating in the conservation program is the payback of the loan and the annual interest:

$$R = c + (1 + r)^t(V_1 - c) \tag{1}$$

In the absence of transaction costs, equation (1) also determines the government costs of conservation. We next plug this equation to an auction model.

We assume that when inviting the bids, the government promises to pay landowners exactly for the bids, thus resorting to a discriminatory reverse auction. This model was initially presented by Latacz-Lohmann and van der Hamsvoort (1997) for agriculture and applied for the first time for forestry by Juutinen and Ollikainen (2010). In this model, the forest landowners' strategies depend on the notion of a maximum acceptable level of the downpayment above which no bids are accepted. Thus, they form their expectations on this bid cap. Let the landowners' expectations about the implicit bid cap be uniformly distributed over the range $[\underline{c}, \overline{c}]$, where the lower bar represents the minimum, and upper bar the maximum expected bid cap, that is, the downpayment. The probability that a landowner's bid is accepted to the conservation program is given by:

$$P(\theta \leq c) = \int_{\underline{c}}^{\overline{c}} f(c)dc = 1 - F(c), \tag{2}$$

where $f(c)$ is the probability density function associated with $F(c)$ and $c$ depends on the requested initial downpayment $c$.

We distinguish between two types of forest landowners, those who maximize only net harvest revenue and those who maximize the sum of net harvest revenue and their valuation of amenity benefits produced by the stand (Amacher et al. 2008). We call the first type a Faustmannian landowner, and the latter type a Hartmanian landowner. The names refer to the two rotation frameworks that characterize these different landowner motives towards forest production (Faustmann 1848; Hartman 1976). Furthermore, Hartmanian landowners' preferences towards amenities differ leading to differing management regimes. These preferences are not, however, visible for the government (see Amacher and Ollikainen (2024) for an analysis of a policy under uncertainty in a rotation framework).

We first develop the expected net payoff and optimal bid for the Hartmanian landowner and derive those for the Faustmannian landowners by setting amenity valuation equal to zero. Let $V_0 + A_0$ denote the net present value of harvests and amenity benefits for Hartmanian landowners when they do not participate in the conservation program and run their optimal management plan. Furthermore, let $V_1 + A_1$ denote these values when they become enrolled in the program. The decision whether or not to participate in the program, depends on the value of the stand, the amenity benefits and the return to the deferrence mechanism. Combining this notation with equation (1) gives the expected net payoff of a risk-neutral landowner from bidding, $J$. It is a product of the revenue from winning the bid and the acceptance probability,

$$J = \left[(V_1 - V_0) + (A_1 - A_0) + c + (1+r)^t \frac{V_1 - c}{x}\right](1 - F(\theta)), \tag{3}$$

The downpayment is the landowner's choice variable, and it is the counterpart to the up-front bid in the ordinary auction models applied to forestry (Juutinen and Ollikainen 2010; Kangas and Ollikainen 2025). Choosing $c$ to maximize Eq. (3) yields:

$$\left(1 - (1+r)^t \left(\frac{1}{x}\right)\right)(1 - F(\theta)) - \left[-V_0 + (A_1 - A_0) + c + (1+r)^t \left(\frac{V_1 - c}{x}\right)\right] f(\theta) = 0 \tag{4}$$

To simplify the notation, define $\Omega = \left(1 - (1+r)^t \left(\frac{1}{x}\right)\right) > 0$. By rearranging and solving equation (4) for $c$ produces an implicit form for the optimal bid, that is, for the requested optimal downpayment for the Hartmanian (H) and Faustmannian (F) landowners:

$$c_H^* = \frac{(1 - F(\theta))}{f(\theta)} + \frac{\left[V_0 - V_1(1+r)^t \left(\frac{1}{x}\right) - (A_1 - A_0)\right]}{\Omega} \tag{5}$$

$$c_F^* = \frac{(1-F(\theta))}{f(\theta)} + \frac{\left[V_0 - V_1(1+r)^t\left(\frac{1}{x}\right)\right]}{\Omega} \qquad (6)$$

Furthermore, in order for the landowner to participate and offer their site to conservation, the following participation constraint must be met:

$$c + \sum_1^t v(1+r)^{-t} > V_0 - (A_1 - A_0), \qquad (7)$$

where $v$ denotes an annual instalment.

Using next the properties of the uniform distribution we develop an explicit solution for both bids:

$$c_H^* = \frac{\bar{c}}{2} + \frac{\left[V_0 - V_1(1+r)^t\left(\frac{1}{x}\right) - (A_1 - A_0)\right]}{2\Omega} \qquad (8)$$

$$c_F^* = \frac{\bar{c}}{2} + \frac{\left[V_0 - V_1(1+r)^t\left(\frac{1}{x}\right)\right]}{2\Omega} \qquad (9)$$

Equations (8) and (9) represent a modification of the traditional up-front bid for forest conservation, where $\bar{c}$ denotes the common belief of the highest acceptable bid. From previous literature, the difference to up-front bids relates to the second term. Recall that for the up-front bid of the Hartmanian landowners, we have $\frac{\bar{c}}{2} + \frac{[(V_0 - V_1) - (A_1 - A_0)]}{2}$, where $\bar{c}$ denotes the common belief of the highest acceptable bid (setting parameters $A_1$ and $A_0$ equal to zero gives the formula for Faustmannian landowners). Thus, the new features in equations (8) and (9) relate to the presence of the initial downpayment (first term) and the interest earnings (the second term defining conservation costs) and the interest lost from the loan $\Omega$. The interest earnings decrease the conservation costs for the landowner in the numerator relative to the up-front bids' formula. In contrast, term $\Omega$ enters the denominator in equations (8) and (9). For the relevant range of interest rates in the auction mechanism, we have that $0 < \Omega < 1$. Thus, it increases the size of the cost term tending to increase the requested downpayment.

Given that equations (8) and (9) represent explicit solution for the requested initial instalment, we can differentiate it to see how the features of the auctioning mechanism and other exogenous parameters impact the optimal solution.

$$\frac{\partial c}{\partial V_0} = \frac{1}{2\Omega} > 0; \frac{\partial c}{\partial V_1} = -\frac{(1+r)^t \frac{1}{x}}{2\Omega} < 0 \tag{10}$$

$$\frac{\partial c}{\partial A_0} = \frac{1}{2\Omega} > 0; \frac{\partial c}{\partial A_1} = -\frac{1}{2\Omega} < 0 \tag{11}$$

$$\frac{\partial c}{\partial r} = \frac{-t(1+r)^{t-1} 2t(V_0 - (A_0 - A_1))}{4\Omega^2} < 0; \frac{\partial c}{\partial \bar{c}} = \frac{1}{2} > 0 \tag{12}$$

The economic interpretation of equations (9) and (10) are straightforward: while a higher net present value of harvesting and amenity benefit valuation from forestry increases the size of the downpayment, rewards from conservation in the form of high conservation payment and private amenity valuation decrease it. A higher interest rate makes giving the loan to the government more profitable and decreases the required downpayment. Finally, an increase in the collective belief on the bid cap increases the required downpayment. These results provide a testable hypothesis for our numerical simulations.

Next, we present the data and formalize the amenity value function.

### 3. Data and specifications of the numerical model

#### 3.1. Data

The data includes in total 400 sites permanently protected in the METSO programme, located in North Karelia in eastern Finland, and in Southwest Finland and Satakunta in western Finland. The data has been collected with field assessments by Parks & Wildlife Finland, which manages state-owned nature reserves in Finland (SAKTI 2019). The sites vary so that they cover all Finnish forest site types (classified based on fertility, see e.g., Tonteri et al. 1990), different tree species compositions and a variety of different age classes (6–230 years[1]). The data includes information on sites' stand age, stand volume, tree species composition and the amount of deadwood. We have two data sets: a set of 100 sites with more detailed information on the stands and a set of 300 sites with less detailed basic information. Appendix A presents the data and Appendix B describes how the data was treated to calculate conservation payments and opportunity costs for all sites. We assume that the area of each site is 10 hectares (an average area of METSO sites). The conservation payment of the program comprises the stand timber value and a fixed 400 € payment for land value. The opportunity costs of landowners, estimated with a stand simulation model MOTTI (Hynynen et al. 2005), were calculated for the more detailed set of 100 sites and generalised to the rest of 300 sites (see Appendix A and B).

---

[1] The data includes some young sapling stands, which may have been selected to METSO due to their connectivity to more valuable stands. We include a wide range of age classes in the data to examine how the different mechanisms select different age classes for conservation.

### 3.2. Amenity benefit valuation: parametric specification

Landowners' attitudes towards conservation range from strong conservation motives to financially oriented forest owners, and to forest owners who are not interested in conservation measures at all (Koskela and Karppinen 2021). Following the theory, we parametrize the conservation-minded Hartmanian landowners or financially driven Faustmannian landowners as follows. We identify Hartmanian landowners in the data by sites that are older than the commercial rotation age (determined based on the recommendations on Best Practices for Sustainable Forest Management in Finland (Tapio 2022)). Faustmannian landowners follow the commercial rotation age (Amacher et al. 2009, Juutinen et al. 2013), and their amenity valuation is zero.

We describe the preferences of Hartmanian landowners towards amenities using the amenity valuation function developed in Juutinen (2005) (see also Juutinen et al. 2013 and Kangas & Ollikainen 2025). We assume that the landowners derive amenity benefits from old-growth stand characteristics. Thus, the valuation is described by a logistic function where the amenity values, A, depend on the stand age ($h$) as follows:

$$A(h) = \frac{1}{\left(\frac{1}{\emptyset K_{max}} + d_0 d_1^h\right)}. \tag{13}$$

Parameters $d_0$ and $d_1$ determine the shape of the sigmoid function ($d_0, d_1 > 0$). They were set to $d_0 = 0.04$ and $d_1 = 0.95$ (the same for all landowners). Parameter $K_{max}$ denotes the cumulative maximum conservation value (€/ha), and we assume that it is 23 500 €/ha without discounting. This value was assessed following Juutinen (2008) for an annual estimate and extending it cumulatively for 50 years. As landowners have varied preferences which depend also on other than stand characteristics (Koskela and Karppinen 2021), parameter ∅ is a random variable covering this variation. It is drawn from a normally distributed set (mean = 1, standard deviation = 0.2). Under this modelling, the size of the amenity benefits valuation under landowner's private harvesting plan, $A_0$ is calculated by setting parameter $h$ in Eq. (13) equal to the optimal harvest age of the stand. Respective value $A_1$ under participation in the conservation program is calculated by setting $h = current\ stand\ age + 50\ years$.

### 3.3. Site selection criteria: ecological value and stand age

We employ two site selection criteria in the analysis: benefit/cost criteria and criteria for ecologically valuable old-growth stands. Both require assessing the ecological value of each site.

We assess the ecological value of each site using use a habitat-based biodiversity metric, ELITE index. It calculates the ecological value based on habitat-specific structural components (Kotiaho et al. 2016; Kangas et al. 2021). The components include the amounts of dead wood, large trees, and broad-leaved trees (in fertile sites) or burnt area (in barren sites). These indicators are similar to the METSO selection criteria.

In the ELITE index, the current state of each component is compared against a predefined reference state. Weights in Eq. (14) reflect the importance of each component to the overall state. Reference values and weights have been determined drawing on expert assessments (Kotiaho et al. 2016). The index value ranges from 0 to 1, where 1 is the reference state (natural state or target state of the ecosystem) and 0 implies that the ecosystem is completely degraded. The index calculates the ecological value ($e$) as follows:

$$e = \prod_{n=1}^{N^k} \left(1 - L_n^k \left(1 - \frac{n_{curr}}{n_{ref}}\right)\right), \tag{14}$$

where $N^k$ is the number of components for a habitat type $k$, $L_n^k$ is the weight for component $n$, and parameters $n_{curr}$ and $n_{ref}$ denote the current and the reference states of component $n$, respectively (Kotiaho et al. 2016). Parameter values for $L_n^k$ and $n_{ref}$ are presented in Appendix B in Table A.2. We did two modifications to the original index due to limited data: the number of large trees was replaced with stand age, and the size of burned forest area in xeric and barren heath forests was assumed to be zero.

Table 1 shows stand age criteria for selecting ecologically valuable old-growth forests. The criteria are determined for the METSO programme (Syrjänen et al. 2016), and they depend on the forest site type and tree species.

**Table 1.** Stand age criteria for different forest site types (Syrjänen et al. 2016)

| Forest site type | Criteria |
|---|---|
| Herb-rich forests | $70^1$, $100^2$ |
| Herb-rich heath forests | $80^1$, $100^2$ |
| Mesic heath forest | $80^1$, $120^2$ |
| Sub-xeric heath forest | 140 |
| Xeric heath forest | 140 |
| Barren heath forest | $140^3$ |

$^1$ Dominated by broad-leaved trees, $^2$ dominated by coniferous trees, $^3$ The criteria (all succession stages close to natural state) not possible to apply due to limited data, so we use stand age instead.

Finally, the site selection is impacted by harvesting. By using a random variable, we determine which stands are harvested during the time period of the analysis. The variable is based on the assumption that in 40 years, all stands passing the commercial rotation age will be harvested.[2]

## 4. Results

In this section, we examine the properties of the deferrence auction for forest conservation. Also, we compare these results to an up-front payment mechanism for forest conservation. To make the solutions comparable, we employ in all calculations conservation budgets that differ annually but have identical net present values over the time horizon.

### 4.1. Functioning and properties of the deferrence mechanism

The main finding of our simulations is that the deferrence mechanism really serves what the theory suggests: on average, the downpayments are 51% of the total conservation payment, the variation being 3-228%. The conservation payment (excluding interest) is on average 7 300 €/ha and the downpayment is 4 010 €/ha. Including the interest, the total conservation payment is on average 8 440 €/ha, and the annual instalment is 467 €/ha.

Figure 1 presents all sites in terms of total conservation payments, downpayments and opportunity costs. The length of the period is ten years, and the interest rate is 3%. Recall, the conservation program defines the rules to calculate the conservation payment for each site, thus making the payment dependent on timber characteristics of the site. The expected bid cap ($\overline{c}$) moderates the bids for downpayments which can be seen from the most expensive sites.

---

[2] This is a rather conservative estimate as it leads to 1–4% being harvested annually, on average 3%. The average harvest rate for mature forests in Finland is 7% (Moilanen & Kotiaho 2020).

Therefore, the more expensive the site, the more useful the instalment mechanism: Competition between landowners brings most benefits with these sites.

With some sites, the bid for the downpayment exceeds the conservation payment because these sites have high opportunity costs which are not completely covered by the conservation payment[3]. The owners of these sites do not participate in conservation as their decision is based on equation (7) and sites with these properties do not fulfil the constraint.

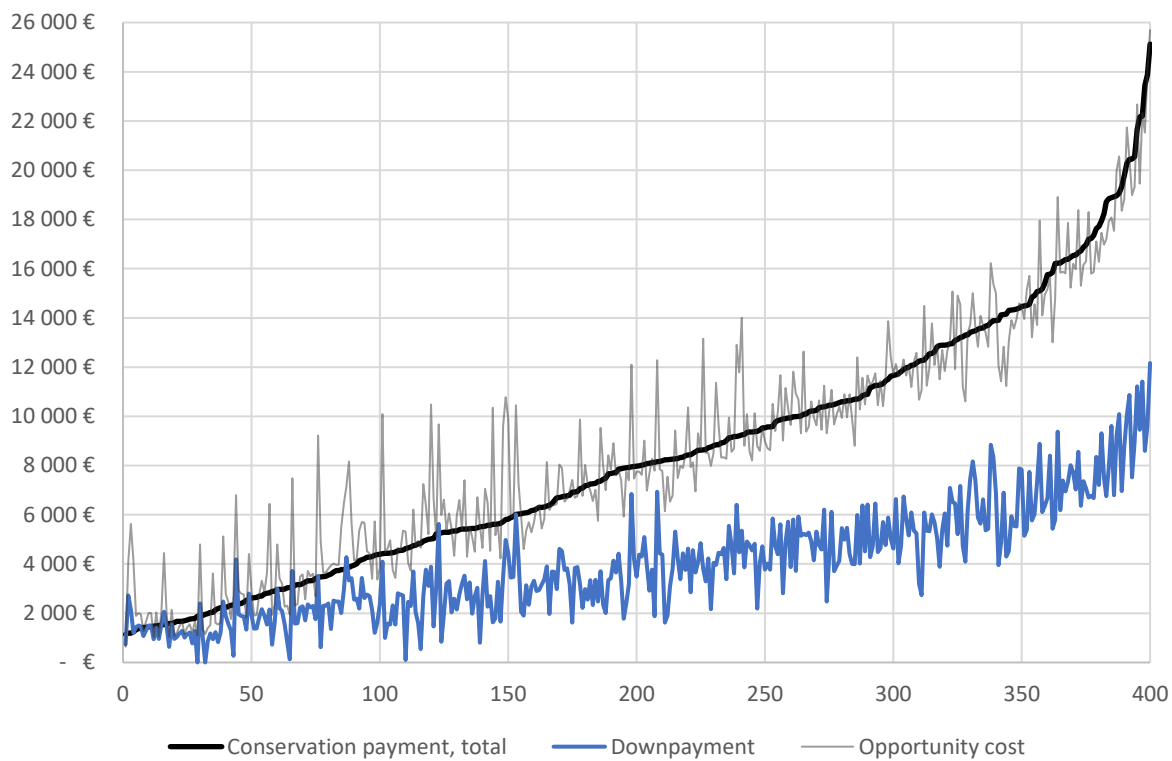

**Fig. 1** All sites in terms of total conservation payments, downpayments and opportunity costs, ranked based on the conservation payment from lowest to highest

If the entire data set were conserved, the downpayments at the time of conservation would cost 1.6 million euros. If, instead, the same area was conserved with up-front payments, the cost would be 2.9 million euros. With the deferred payments, the subsequent instalments would cost approximately 178 000 euros per year for ten years, and 1.5 million euros in total in net present value (downpayments and instalments 3.1 million euros in total).

---

[3] The payment covers the value of the stand and a fixed payment for land. If the bare land value (net present value of future rotations) is very high but the current stand volume low, the conservation payment does not cover the opportunity cost.

Table 2 presents the supply of sites for conservation under the deferred payment mechanism. Given that the interest rate plays an important role for the requested downpayments (cf., Eq. 12), we consider landowners' incentives under three different interest rates.

**Table 2.** Supply of sites under the deferred payment mechanism, a 10-year payment period. Different options for the interest rate.

|  | 3% | 2% | 4% |
|---|---|---|---|
| Area, ha | 2 470 | 2 240 | 2 750 |
| BD index, sum | 748 | 679 | 823 |
| Stand age, avg., years | 110 | 112 | 107 |
| Downpayment, avg., €/ha | 3 770 | 3 540 | 3 890 |
| Instalment, avg., €/ha/year | 546 | 488 | 594 |
| Total payment, NPV, avg., €/ha | 8 430 | 7 700 | 8 960 |

Table 2 shows that a higher interest rate increases landowners' participation because of higher interest earning for the loan. With a 10-year payment schedule and 3% interest rate, 2470 hectares is offered to the programme, which is 62% of the total area. Thus, the offered area of sites increases in an increasing fashion with the interest rate.

For comparison to the deferrence mechanism, Table 3 presents the supply incentives under up-front payments.

**Table 3.** Supply of sites under up-front payments.

| Area, ha | 2 150 |
|---|---|
| BD index, sum | 653 |
| Stand age, avg., years | 113 |
| Up-front payment, avg., €/ha | 7 500 |

With the up-front payment mechanism, 2150 hectares is offered to the programme, which is 54% of the total area. Thus, landowners participate more with the deferred payment mechanism. Paying interest for the loan increases the landowners' income from conservation so that more landowners are willing to participate (recall equation (7)).

## 4.2. Site selection results: benefit/cost ranking

Next, we analyse the site selection results with both payment mechanisms. Recall, the government conservation budgets are defined so that the net present value of the budgets under both mechanisms are equal. The annual cost burden, however, differs with the mechanisms. The up-front payment mechanism spends the same annual budget for the whole period whereas the deferred payment mechanism uses a bigger initial budget at year 0 for the downpayments and smaller subsequent budgets for the instalments. The selection employs benefit/cost criteria.

Figure 2 shows the cost burden for both mechanisms, depicting annual budgets and cumulative budgets.

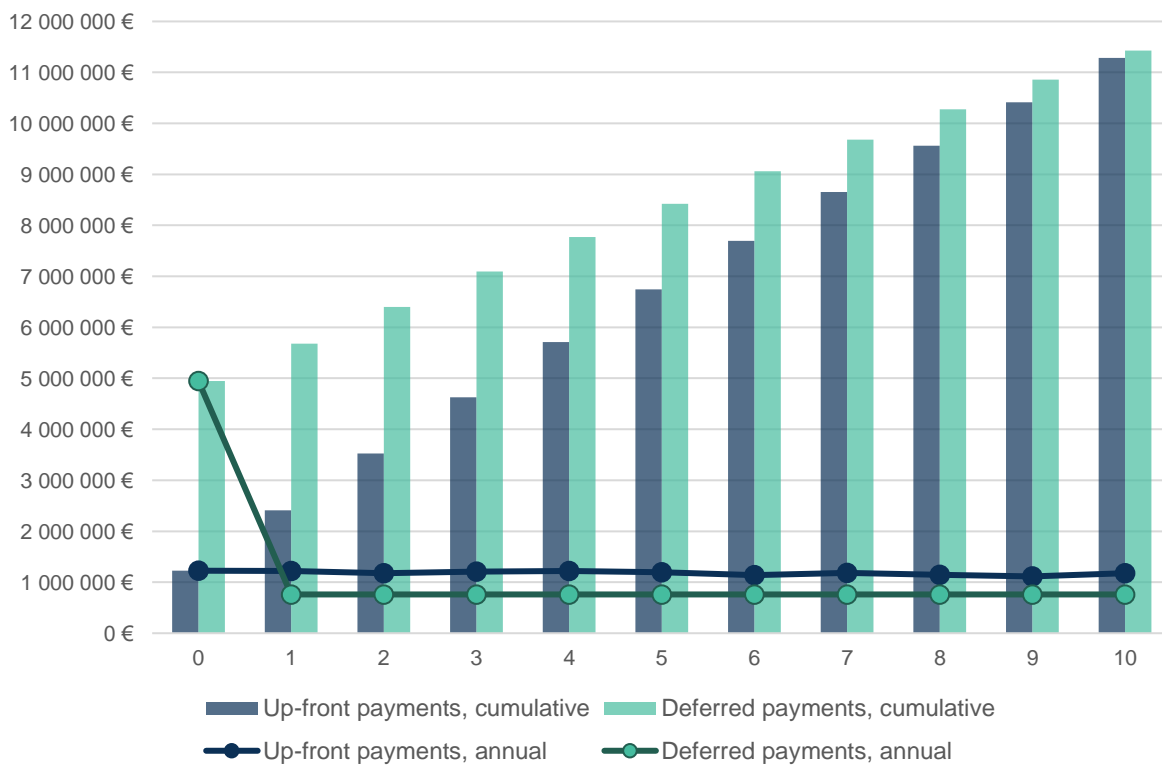

**Fig. 2** The cost burden of conservation under both payment mechanisms: Lines with markers depict annual conservation costs (y-axis on the left) and columns depict cumulative conservation costs (y-axis on the right)

Figure 2 shows clearly that difference between the two mechanisms lies in how the conservation costs are distributed over time. Lines depicting the annual costs differ but in the end of the period, the cumulative costs are approximately the same. In absolute values, the deferred payment causes slightly less costs (10.2 million euros) than the up-front mechanism (10.7 million euros).

Figure 3 in turn shows the site selection results in terms of land areas conserved under both payment mechanisms.

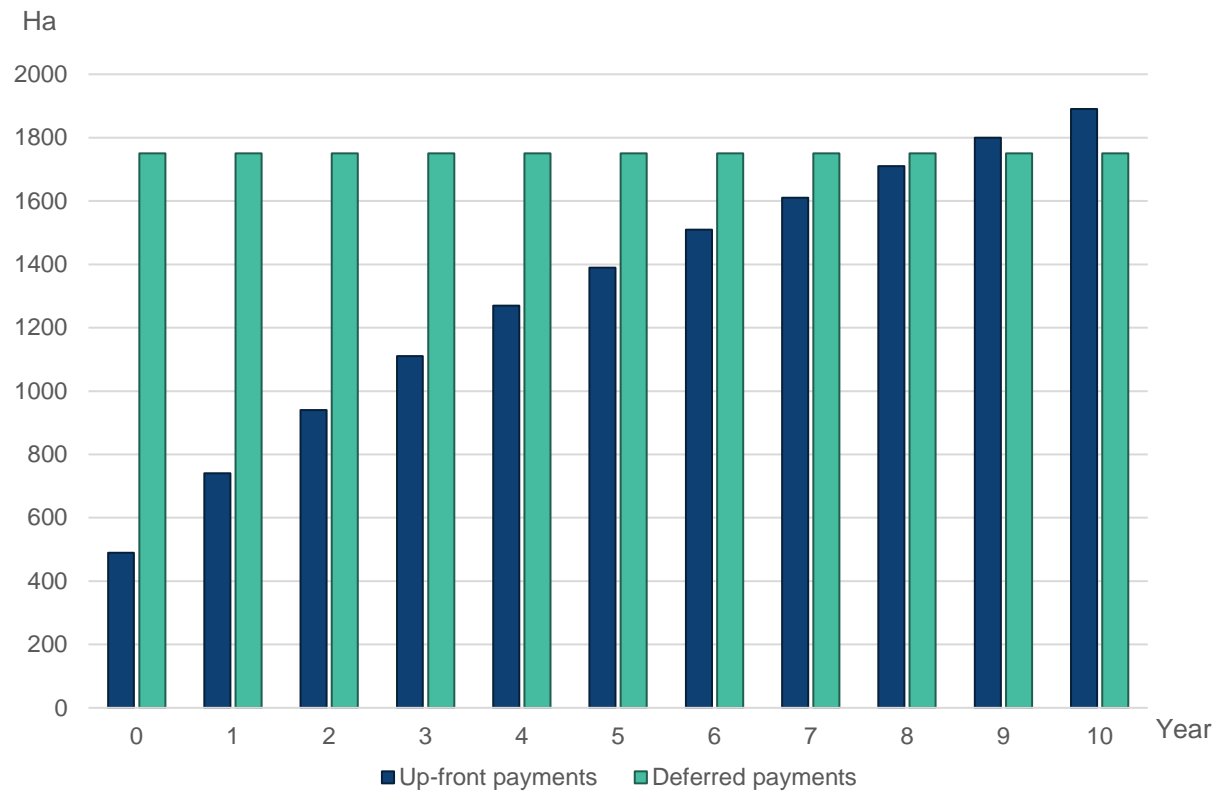

**Fig. 3** Cumulative conservation areas with different payment schedules

From Figure 3, the up-front payment mechanism conserves a slightly larger land area but with the deferred payment mechanism, the whole area is conserved in year 0. With the up-front payments, the conservation area is expanded gradually. This means that some sites are harvested before they can be conserved.

Table 4 condenses information of the selected sites under both payment mechanism. We report the cost items, biodiversity properties, and costs from harvested valuable stands (up-front-payments). Also, we employ valuation studies to determine the present value of net benefits from both mechanisms. The last row reports the ex post net benefits as a figure that includes lost benefits due to harvesting during the time span of the program.

**Table 4**. Results with the deferred payment mechanism (10 year-payment schedule, 3% interest) and up-front payments, under benefit/cost ranking.

| **Deferred payments** | | **Up-front payments** | |
|---|---|---|---|
| Initial budget, € | 5 000 000 | Annual budget, € | 1 250 000 |
| Cost of instalments, €/year | 760 000 | Cost of instalments, €/year | 0 |
| Area, ha | 1 750 | Area, ha | 1 890 |
| Harvested sites, ha | 0 | Harvested sites, ha | 160 |
| BD index sum | 547 | BD index, sum | 581 |
| Stand age, avg., years | 110 | Stand age, avg., years | 112 |
| Downpayments, avg., €/ha | 2 830 | Payments, avg., €/ha | 6 880 |
| Instalments, avg., €/ha/year | 434 | Instalments, avg., €/ha/year | 0 |
| Costs, total (NPV), € | 11 427 100 | Costs, total (NPV), € | 11 286 500 |
| Benefits, total (NPV), € | 10 465 000 | Benefits, total (NPV), € | 10 280 700 |
| Lost benefits in harvests, € | - | Lost benefits in harvests, € | - 956 800 |
| Ex post net benefits, € | 10 465 000 | Ex post net benefits, € | 9 328 900 |

Due to the additional cost from interest paid to deferred payments, it conserves 7% less land compared to the up-front payment mechanism under the same net present value budget[4]. The up-front payments are on average 6900 €/ha. The average downpayment of the selected sites is 1750 € per hectare, and the instalments 430 € per hectare per year. For comparison, the annual conservation payment for 20-year temporary conservation agreements in METSO was 102 €/ha/year in 2023 (harvests are not allowed, covers lost revenue only from the period of the agreement, 10 years) (Koskela et al. 2024). In Mäntymaa et al. (2018) survey for landowners, the compensation claim for a 5-year forest landscape conservation scheme was approximately 330 € per hectare per year (soft harvests allowed).

We estimate the benefits of conservation based on Lehtonen et al. (2003), 5 980 €/ha (in current value of money). The benefits are calculated in net present value (including the value of the site in the calculation at the year it is conserved). Therefore, the deferred payments lead to higher total benefit as the whole area is conserved in year 0 whereas up-front payments conserve sites gradually. In the case of up-front payments, 160 hectares are harvested during the 11-year period and are no longer offered for conservation at the time they would be selected. Stand age of the logged sites is on average 125 years and the lost conservation benefit is approximately 957 000 €. We take the lost benefits of harvests into account when calculating the ex post net benefit value.

---

[4] With a lower budget, the difference between the mechanisms is approximately the same in terms of land area and monetary benefits. With a bigger budget, the set of supplied sites runs out with the up-front payment mechanism before the conservation budget is exhausted. As the supply of sites is higher under the deferred payment mechanism, it leads to a better outcome both in terms of land area conserved and the monetary benefit of conservation.

Table 5 and 6 show as a sensitivity analysis the results if the interest rate paid for the instalments is decreased to 2% or increased to 4%.

**Table 5**. Results with the deferred payment mechanism (10 year-payment schedule, 2% interest) and up-front payments, benefit/cost ranking.

| Deferred payments | | Up-front payments | |
|---|---|---|---|
| Initial budget, € | 5 000 000 | Annual budget, € | 1 200 000 |
| Cost of instalments, €/year | 702 600 | Cost of instalments, €/year | 0 |
| Area, ha | 1 730 | Area, ha | 1 860 |
| Harvested sites, ha | 0 | Harvested sites, ha | 160 |
| BD index sum | 542 | BD index, sum | 575 |
| Stand age, avg., years | 110 | Stand age, avg., years | 111 |
| Downpayments, avg., €/ha | 2 870 | Payments, avg., €/ha | 6 790 |
| Instalments, avg., €/ha/year | 406 | Instalments, avg., €/ha/year | 0 |
| Costs, total (NPV), € | 10 965 600 | Costs, total (NPV), € | 10 959 400 |
| Benefits, total (NPV), € | 10 345 400 | Benefits, total (NPV), € | 10 109 600 |
| Lost benefits in harvests, € | - | Lost benefits in harvests, € | - 956 800 |
| Ex post net benefits, € | 10 345 400 | Ex post net benefits, € | 9 152 800 |

Supply of sites to the deferred payment mechanism reduces (Table 2) when the interest rate is decreased which means that the conservation outcome is slightly worse than with a 3% interest rate. The cost of annual instalments decreases due to the lower interest rate. The site selection results for the up-front payments change from Table 4 because the conservation budget is moderated to match the net present value of the budget of the deferred payments. The deferred payment mechanism conserves 7% less land area in this case, as well, so the relative difference does not change. The net present value of benefit is still higher with the deferred payments, reflecting the advantage of conserving the whole area immediately.

Table 6. Results with the deferred payment mechanism (10 year-payment schedule, 4% interest) and up-front payments, benefit/cost ranking.

| Deferred payments | | Up-front payments | |
|---|---|---|---|
| Initial budget, € | 5 000 000 | Annual budget, € | 1 305 000 |
| Cost of instalments, €/year | 826 100 | Cost of instalments, €/year | 0 |
| Area, ha | 1 780 | Area, ha | 1 950 |
| Harvested sites, ha | 0 | Harvested sites, ha | 150 |
| BD index sum | 560 | BD index, sum | 598 |
| Stand age, avg., years | 107 | Stand age, avg., years | 112 |
| Downpayments, avg., €/ha | 2 800 | Payments, avg., €/ha | 7 090 |
| Instalments, avg., €/ha/year | 464 | Instalments, avg., €/ha/year | 0 |
| Costs, total (NPV), € | 12 027 100 | Costs, total (NPV), € | 12 005 400 |
| Benefits, total (NPV), € | 10 644 400 | Benefits, total (NPV), € | 10 613 200 |
| Lost benefits in harvests, € | - | Lost benefits in harvests, € | -897 000 |
| Ex post net benefits, € | 10 644 400 | Ex post net benefits, € | 9 716 200 |

When the incentives of the deferred payment mechanism increase with the increased interest rate, the conservation outcome is slightly better. The differences to tables 4 and 3 are still moderate. Again, up-front payments lead to larger land areas conserved but deferred payments provide higher ex post net benefits.

Results with a 20-year payment schedule and sensitivity analysis on the common expectation on $\overline{c}$ is presented in Appendix C. Lengthening the payment period decreases landowners' incentives to participate and considerably reduces supply of sites to the program, which worsens the conservation outcomes with the deferred payment mechanism. Increasing the expectation on $\overline{c}$, the highest downpayment still winning the auction, increases downpayments and slightly increases supply, and vice versa if $\overline{c}$ is decreased. However, the difference between the mechanisms remains approximately the same in terms of land area (the deferred payment leads to 7-9% smaller area) and monetary benefits (the deferred payment leads to higher ex post net benefit).

### 4.3. Site selection results: selection of old-growth forests

Next, we change the selection criteria to target ecologically valuable old-growth stands. The criteria for different forest site types were presented in Table 1. In this case, the conservation budget is not exhausted because all offered old-growth sites are conserved before the budget is exhausted. Therefore, this section differs from the previous one and we focus on the conservation outcome for the case where all offered old-growth sites are conserved with both mechanisms (but conservation budget is not exhausted). Table 8 presents the results.

Table 7. Site selection results with stand age criteria.

| Deferred payments | | Up-front payments | |
|---|---|---|---|
| Initial budget, € | 3 647 400 | Annual budget, € | 900 000 |
| Cost of instalments, €/year | 613 200 | Cost of instalments, €/year | 0 |
| Area, ha | 900 | Area, ha | 750 |
| Harvested sites, ha | 0 | Harvested sites, ha | 70 |
| BD index sum | 294 | BD index, sum | 247 |
| Stand age, avg., years | 147 | Stand age, avg., years | 148 |
| Downpayments, avg., €/ha | 4 050 | Payments, avg., €/ha | 8 620 |
| Instalments, avg., €/ha/year | 681 | Instalments, avg., €/ha/year | 0 |
| Costs, total (NPV), € | 8 877 900 | Costs, total (NPV), € | 5 864 700 |
| Benefits, total (NPV), € | 5 832 000 | Benefits, total (NPV), € | 4 190 500 |
| Lost benefits in harvests, € | - | Lost benefits in harvests, € | -418 600 |
| Ex post net benefits, € | 5 832 000 | Ex post net benefits, € | 3 771 900 |

More ecologically valuable old-growth stands are supplied and therefore conserved with the deferred payment mechanism. The whole area is conserved at once so there is no harvesting risk. Because conserving these sites is more expensive than conserving younger stands, the downpayments are now higher than under benefit/cost ranking, on average 4 050 €/ha, and the annual instalment 680 €/ha. Same figures with benefit/cost ranking were 2 830 €/ha and 430 €/ha (Table 3). The up-front conservation payment is on average 8 620 €/ha, and 6 880 €/ha with benefit-cost ranking.

Figure 4 demonstrates the differences in conserving the ecologically valuable old-growth stands. The deferred payment mechanism conserves the whole area at year 0: 900 hectares of old-growth stands. The up-front payment mechanism conserves the stands gradually until year 7 when all ecologically valuable old-growth sites supplied to the scheme have been conserved. Before the conservation decision, 70 hectares of ecologically valuable old-growth stands are harvested. The lost conservation benefit is 418 600 euros.

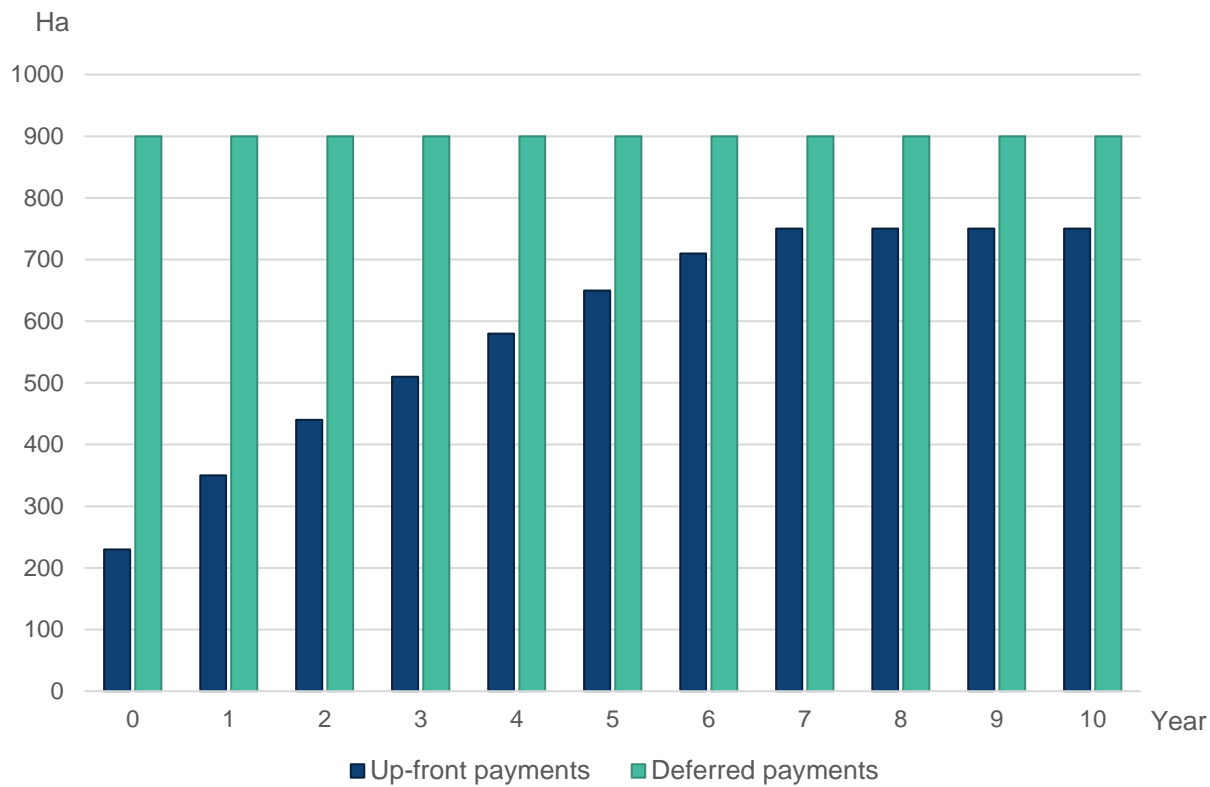

**Fig. 4** Cumulative conservation areas with different payment schedules and ecologically valuable old-growth stand selection

Our analysis brings input in estimating the cost of meeting the EU conservation targets, i.e., that all remaining old-growth stands and primary forest stands must be protected. We employ the average values calculated above for old-growth stands. There are several different estimations concerning the area of remaining primary and old-growth forests in Finland. Syrjänen et al. (2024) estimate the area of old-growth forests in private landowners' forests to be at most 54 000 hectares. Conserving this area with the deferred payment mechanism would cost approximately 219 million euros in downpayments and 37 million euros in annual instalments, in total 533 million euros in net present value. Using up-front payments, the conservation cost would total 456 million euros in net present value (annual budget of 48 million euros), assuming that the total area of 54 000 hectares is conserved evenly, 4 910 hectares per year.[5] This calculation does not take into account the fact that the deferred payment mechanism provided stronger incentives to offer sites to conservation, and that with the up-front payment mechanism, some of the sites may be harvested before they can be conserved. Assuming the same rate of harvests than in

---

[5] In absolute values, the cost from the deferred payment mechanism is 587 million euros and 527 million euros from the up-front mechanism.

Table 8 (9,3% in 10 years), 5000 hectares out of the total 54 000 hectares of old-growth forests would have been lost due to harvesting before protection.

## 5. Discussion and conclusions

The EU biodiversity strategy requires conserving strictly 10% of land and sea areas, including all primary and old-growth forests, by 2030. This requires promptly expanding the protected area network. As an adaption from consumer markets, we introduced a deferred payment mechanism with an auction for downpayments as a novel instrument to conserve large land areas in a quick schedule. The promise of the mechanism is to facilitate a quick conservation of larger land areas relative to conservation with ordinary up-front payments.

We developed a formal framework for the analysis of a deference auction and provided numerical simulations to assess its properties and to compare it to a traditional up-front payment scheme. We demonstrated that the requested initial downpayment works well as the bid in the deferrence mechanism for the given rules of the scheme. The rules fix the total value of the site (conservation payment) and the interest earnings to the loan that the landowner gives to the government. The numerical simulations demonstrated that the deferrence mechanism has the desired property: it facilitates a quick conservation of stands and thereby minimizes the loss of ecologically valuable sites caused by harvesting risks.

We found that the mechanism is sensitive to key parameters, especially to the length of the lending period and the interest rate. The analysis suggests that keeping the lending period no longer than 10 years and paying a 3% interest rate provides a compromise that works rather well and outperforms the up-front mechanisms in monetary terms in every case, as it prevents the loss of valuable stands and yields higher net benefits to the society. In terms of conserved land area, the up-front payment mechanism conserves more but as it includes harvesting risks, the total impact of conservation is worse than with the deferred payment mechanism. When the site selection is made drawing on old-growth stand criteria, the deferred payment mechanism yields a better outcome both in terms of land area and monetary terms as it provides better incentives for landowners to offer old-growth forests for conservation.

Based on theoretical and simulation grounds, the deferrence mechanism performs very well. Before adopting this mechanism in an actual policy, a third step would be necessary. The society should arrange a pilot to test how forest landowners would behave in the experiment and what institutional features would perform best in tailoring the deferrence mechanism as an actual policy tool for conservation. Its promise for cost-effective implementation of the EU's biodiversity strategy is good.

**Acknowledgements**

We thank Metsähallitus for the permission to use the data, and special thanks to Elisa Pääkkö for your help with the data gathering.